\documentclass[aps,prb,amsmath,amssymb,reprint,superscriptaddress,floatfix]{revtex4-2}

\usepackage{graphicx}
\usepackage{xcolor}
\usepackage{color,soul}

\begin{document}

\title{Stability and reactivity of double icosahedron Ag$_{17}$M$_2$ (M=Ni, Cu, Zn) clusters}

\author{P. L. Rodr\'iguez-Kessler}
\email{plkessler@cio.mx}
\affiliation{Centro de Investigaciones en \'Optica A.C., Loma del Bosque 115, Lomas del Campestre, Leon, 37150, Guanajuato, Mexico}

\date{\today}

\begin{abstract}
Herein, the structure and stability of double icosahedron Ag$_{17}$M$_2$ (M = Ni, Cu, Zn) clusters are investigated using density functional theory (DFT) computations. The results indicate that the clusters favor endohedral configurations in the doublet state, as confirmed with four different functionals: BP86, PBE0, B3PW91, and TPSSh. Additionally, the doped clusters exhibit higher ionization energies and electronegativities compared those of the bare Ag$_{19}$ cluster. After doping, the ELF function increases at the Ag sites, which reveals important implications for catalysis.
\end{abstract}


\maketitle

\section{Introduction}

Nanoclusters, particularly those composed of noble metals like silver, have drawn considerable attention due to their unique structural, electronic, and catalytic properties. The arrangement of atoms within these clusters often leads to behaviors that deviate from those of bulk materials, making them ideal candidates for a variety of technological applications. Among these, the double icosahedron structure, characterized by a distinct icosahedral geometry with enhanced stability, is a particularly interesting motif in the study of metallic clusters.\cite{Baletto_2019,JIANG2015156,C3CP55144J} Moreover, clusters of coinage and noble metals have demonstrated enormous potential for electrocatalysis. While noble-metal clusters demonstrate remarkably high catalytic activity in facilitating the four-electron reduction of oxygen molecules, transition metals suffer from corrosion issues under acidic conditions.\cite{Yamamoto2009-ne} Silver and its alloys exhibit notable stability and high activity for the oxygen reduction reaction (ORR) under alkaline conditions, making them attractive, low-cost alternatives to noble-metal catalysts.\cite{C5CY02270C} Previous DFT studies have shown that core-shell Ag nanoclusters, especially those with noble-metal cores, are promising catalysts for ORR.\cite{10.1063/1.4972579,RODRIGUEZCARRERA2024122301} The development of noble-metal-free catalysts, such as Ag-based nanoclusters and alloys, is crucial for expanding the range of catalysts suitable for alkaline fuel cell applications. Moreover, small coinage metal clusters showed promising properties for the hydrogen evolution reaction.\cite{YANG20213484} In this preprint, we anticipate the stability and reactivity of double icosahedral Ag$_{17}$M$_2$ clusters, where M represents the transition metals Ni, Cu, and Zn. The results show that the incorporation of two M atoms into the silver framework introduces significant modifications to their stability and reactivity. The structural stability, bonding interactions, and energetic preferences of these systems are investigated by means of density functional theory (DFT). The findings of this study provide crucial insights into the effect of metal substitution on the stability and chemical reactivity of Ag$_{17}$M$_2$ clusters, highlighting their potential for catalysis and other applications that require enhanced nanomaterial reactivity.

\section{Computational Details}

Calculations performed in this work are carried out by using density functional theory (DFT) as implemented in the ORCA 6.0.0 code.\cite{10.1063/5.0004608} The exchange and correlation energies are addressed by the PBE0 functional in conjunction with the Def2-ECP\cite{B508541A} and auxiliary def2/J\cite{B515623H} basis sets, where ECP stands for effective core potential.\cite{Andrae1990,10.1063/1.452110} Atomic positions are self-consistently relaxed through a Quasi-Newton method employing the BFGS algorithm. The SCF convergence criterion is set to TightSCF in the input file. This results in geometry optimization settings of 1.0e$^{-08}$ Eh for total energy change and 2.5e$^{-11}$ Eh for the one-electron integrals. The  Van  der  Waals  interactions  are  included in the exchange-correlation functionals with empirical dispersion corrections of Grimme DFT-D3(BJ). The electron localization function (ELF) was computed and analyzed using Multiwfn.\cite{https://doi.org/10.1002/jcc.22885} 

Theoretical descriptors such as hardness ($\eta$), electronegativity ($\chi$), and electrophilicity index ($\omega$), can be derived from the Koopman's theorem,\cite{KOOPMANS1934104} and are used to evaluate the chemical stability of Ag$_{17}$M$_2$ clusters,\cite{Pearson2005} which can be obtained by using the following equations:

\begin{equation}\eta=\frac{I-A}{2},\end{equation}

\begin{equation}\chi=\frac{I+A}{2},\end{equation}

\begin{equation}\omega=\frac{\chi^2}{2\eta},\end{equation}

where I and A are the vertical ionization energy and vertical electron affinity parameters, which are calculated following previous works.\cite{RODRIGUEZKESSLER2023121620,RODRIGUEZKESSLER2023116538,RODRIGUEZKESSLER2024122062,RODRIGUEZKESSLER2020155897,D1CP00379H}

\section{Results}

\begin{figure}[ht]
  \vspace{.5cm}
\begin{tabular}{cc}
\resizebox*{0.20\textwidth}{!}{\includegraphics{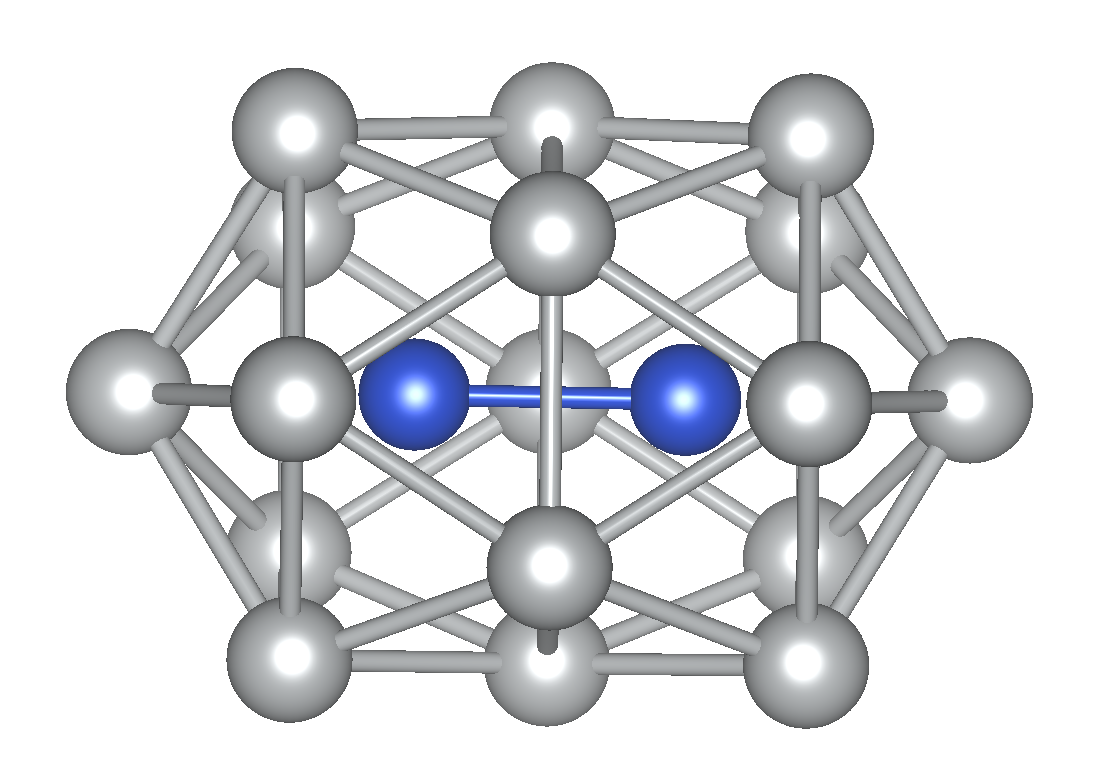}} &
\resizebox*{0.20\textwidth}{!}{\includegraphics{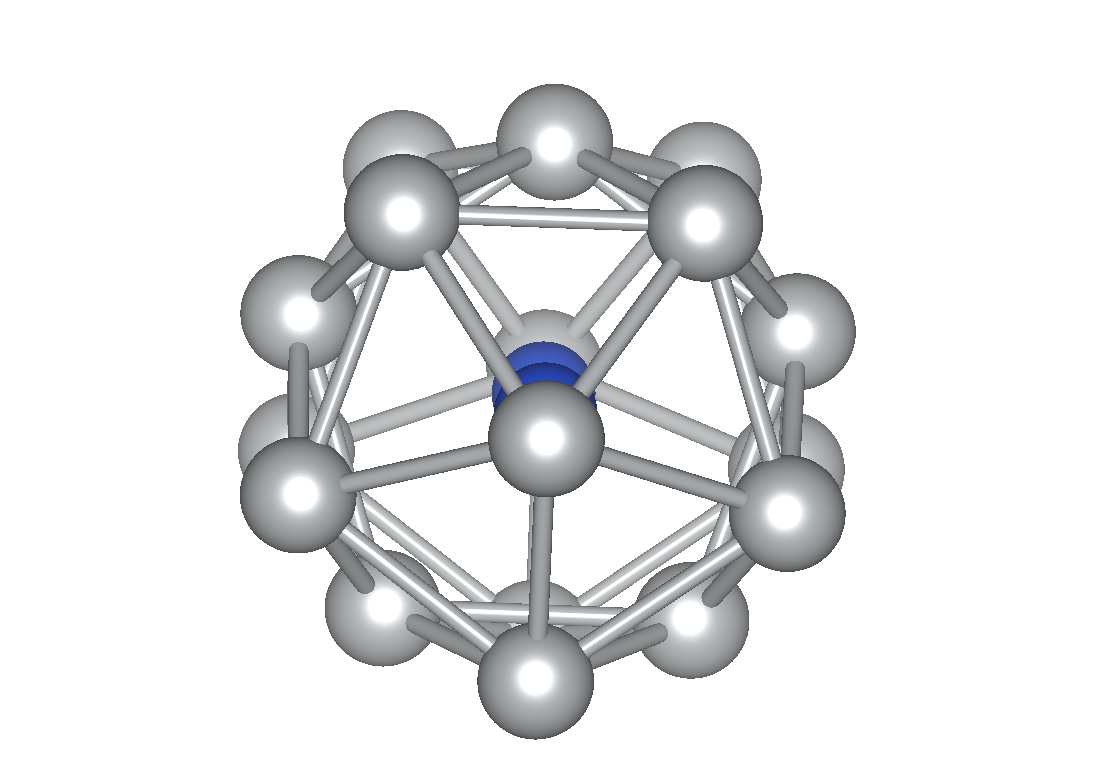}}   \\
     {\bf (a) }  &
     {\bf side-view}  \\
  \resizebox*{0.19\textwidth}{!}{\includegraphics{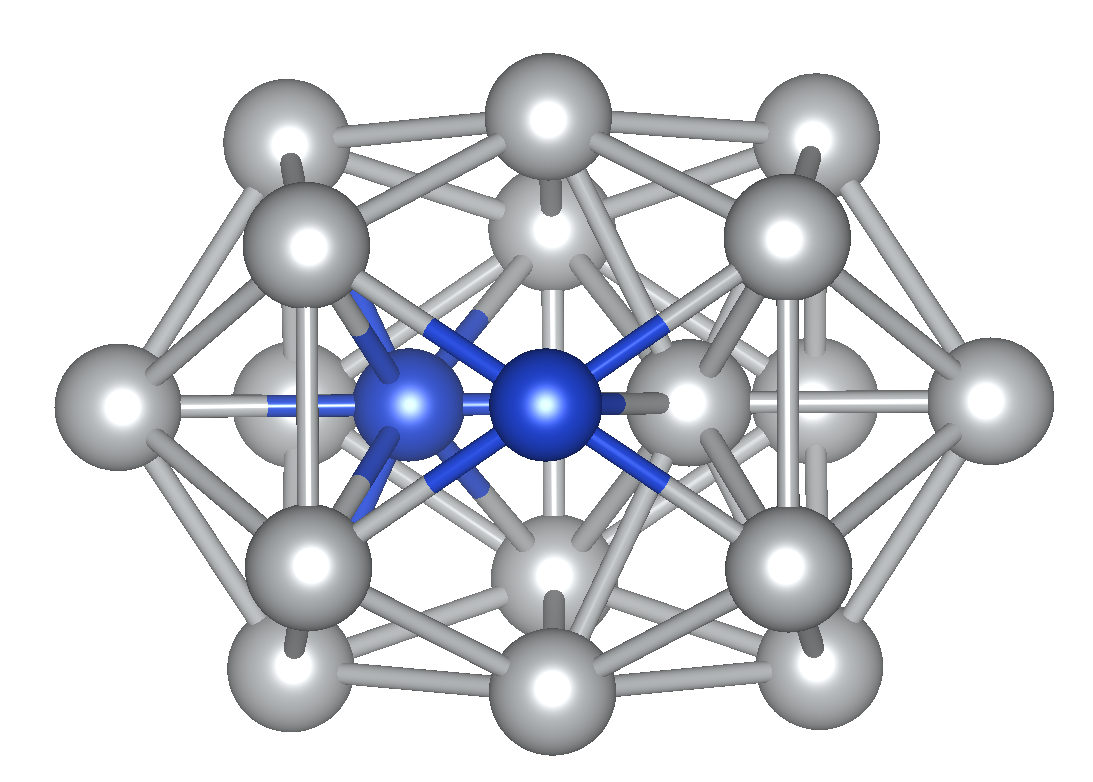}} &
\resizebox*{0.19\textwidth}{!}{\includegraphics{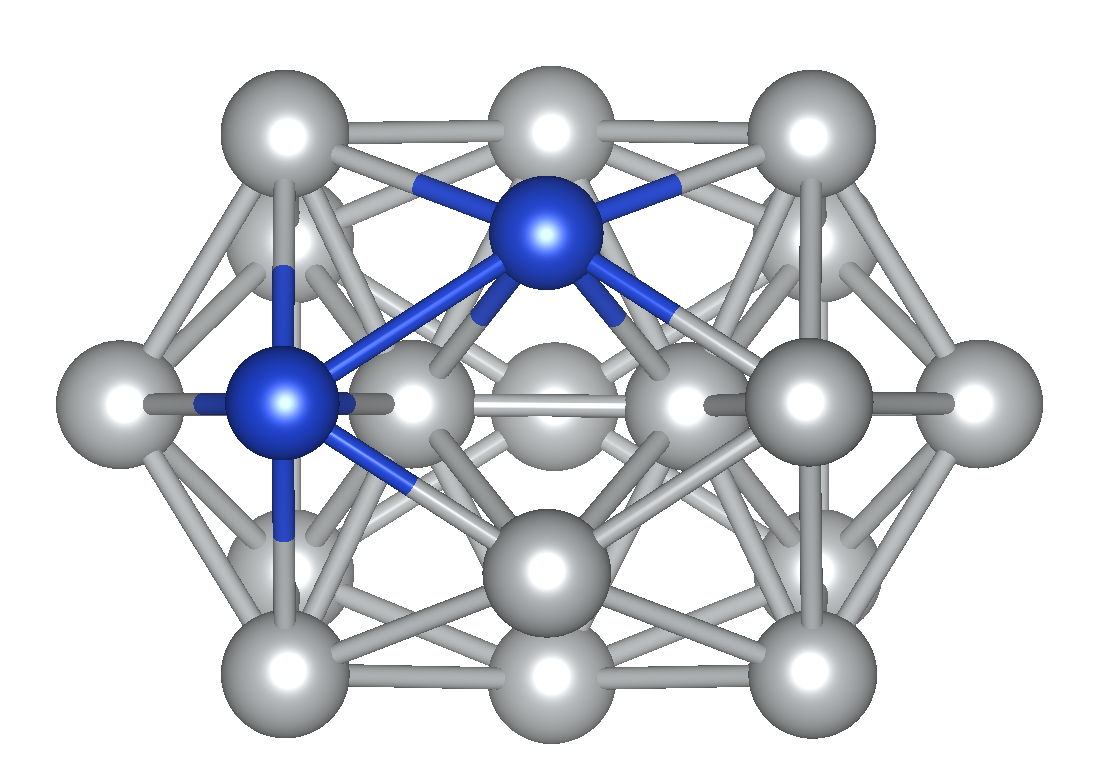}}   \\
   {\bf (b)}  &
   {\bf (c)}  \\
       \end{tabular}
       \caption{\label{figure1}Double icosahedral structure model for representing Ag$_{17}$M$_2$ clusters. Three different chemical orderings (a, b, c) are used to represent the endohedral and exohedral configurations.}
\end{figure}

 The Ag$_{17}$M$_2$ (M=Ni, Cu, Zn) clusters investigated in this work are calculated by using a double icosahedron structure model, as shown in Figure~\ref{figure1}. The results show that the clusters favor the endohedral doping over the exohedral configurations, which is also observed for smaller icosahedral clusters.\cite{RODRIGUEZKESSLER2024141588} Interestingly for Ag$_{17}$Zn$_2$ cluster the endohedral configuration is also more favorable than exohedral configurations (Figures~\ref{figure1}b,~\ref{figure1}c), although the exohedral configuration in the 13-atom icosahedral cluster (Ag$_{12}$Zn) is more stable.\cite{RODRIGUEZKESSLER2024141588} Moreover, the low-spin state (doublet) is favored over the quartet state for these clusters, as confirmed by four different functionals (see Table~\ref{table1}).   

\begin{table}[ht]
\caption{\label{table1}Relative energies (in eV) of the lowest energy structures of Ag$_{17}$M$_2$ (Ni, Cu, Zn) clusters computed at different DFT levels. Representative values on the spin multiplicity (S$_M$) are given.}
\begin{tabular}{p{1.7cm}p{1.0cm}p{1.3cm}p{1.3cm}p{1.5cm}p{1.0cm}}
{Label}  & S$_M$ & \small{BP86} & \small{PBE0} & \small{B3PW91} & \small{TPSSh} \\
\hline
\textbf{Ag$_{17}$Ni$_2$} & 2 & 0.00 & 0.00 & 0.00 & 0.00\\
\textbf{Ag$_{17}$Cu$_2$} & 2 & 0.00 & 0.00 & 0.00 & 0.00\\
\textbf{Ag$_{17}$Zn$_2$} & 2 & 0.00 & 0.00 & 0.00 & 0.00\\
\\
\textbf{Ag$_{17}$Ni$_2$} & 4 & 0.30 & 0.25 & 0.07 & 0.28 \\
\textbf{Ag$_{17}$Cu$_2$} & 4 & 0.52 & 0.43 & 0.45 & 0.47\\
\textbf{Ag$_{17}$Zn$_2$} & 4 & 0.86 & 0.80 & 0.83 & 0.75\\
\hline
\end{tabular}
\end{table}

\begin{figure}[ht]
\begin{center}
\scriptsize
\includegraphics[scale=0.32]{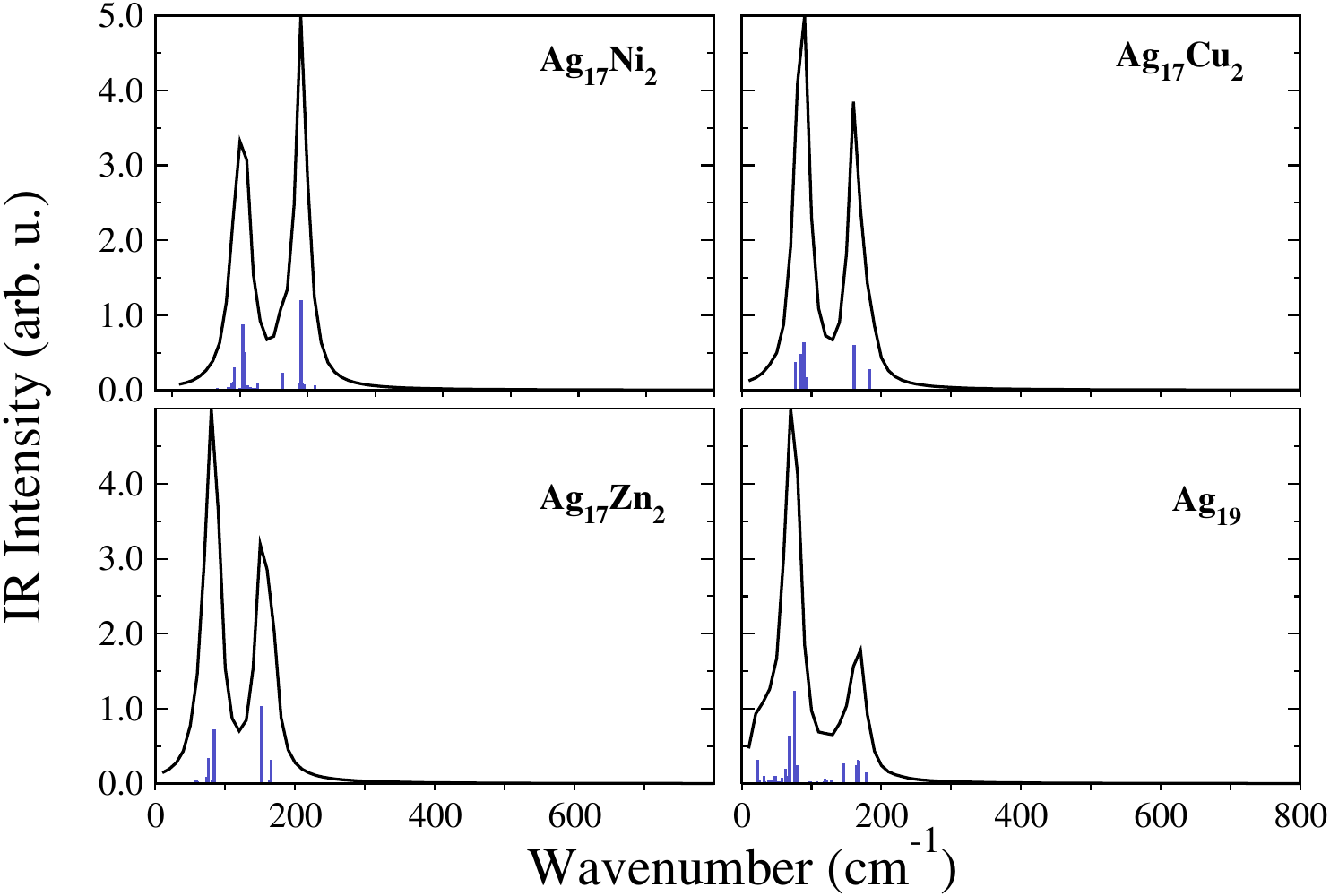}
\caption{\label{figure_IR}IR spectra for Ag$_{17}$M$_2$ (M = Ni, Cu, Zn) clusters obtained at the PBE0/Def2-ECP level. The bare Ag${19}$ clusters is shown for comparison.}
\end{center}
\end{figure}

To identify the fingerprints of the clusters, we have calculated their infrared (IR) spectra, which will serve as a reference for future experimental studies.\cite{2024structuresstabilitiesb7cr2clusters,2024revisitingglobalminimumstructure} All the clusters show a similar IR spectrum with two main peaks, with the intensity depending on the composition, for example, Ag$_{17}$Zn$_2$ and Ag$_{19}$ show the first peak as more pronounced than the second one (Figure~\ref{figure_IR}). The characteristic peak for Ag$_{17}$M$_2$ (M = Ni, Cu, Zn) is found at 190.56, 88.81, and 151.31 cm$^{-1}$, while their vibrational frequency ranges are 46.90–210.92, 39.17–234.72, and 37.12–217.94 cm$^{-1}$, indicating a narrow range for their vibrational spectra. For comparison, we have evaluated the bare Ag$_{19}$ cluster, which shows the main vibrational mode at 75.82 cm$^{-1}$, within a range of 22.11–225.92 cm$^{-1}$.

\begin{figure}[ht]
\caption{\label{figura4}The electron localization function (ELF = 0.7) for a) the Ag$_{17}$Ni$_2$ cluster and b) the Ag$_{19}$ cluster.}
 \resizebox*{0.40\textwidth}{!}{\includegraphics{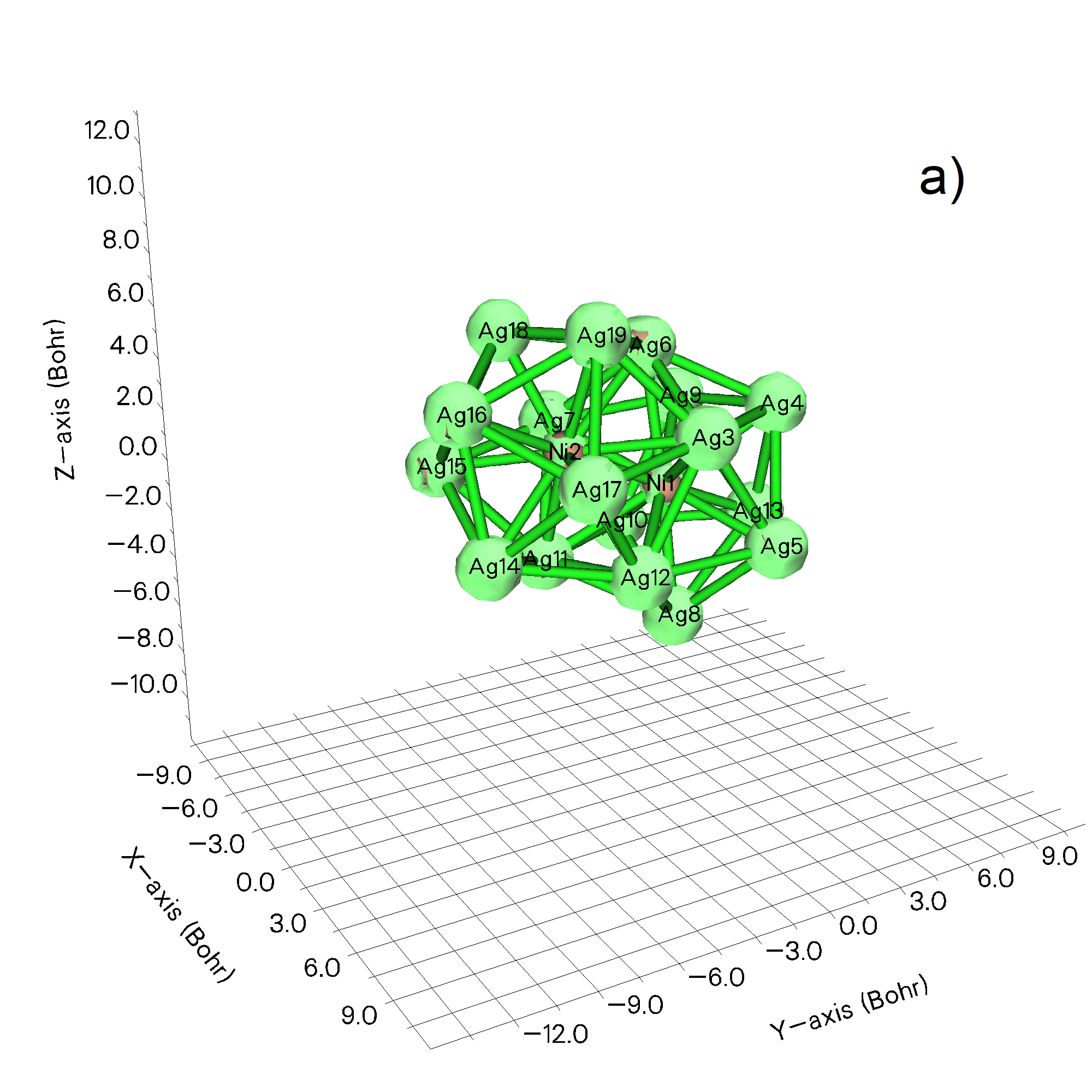}} 
\resizebox*{0.40\textwidth}{!}{\includegraphics{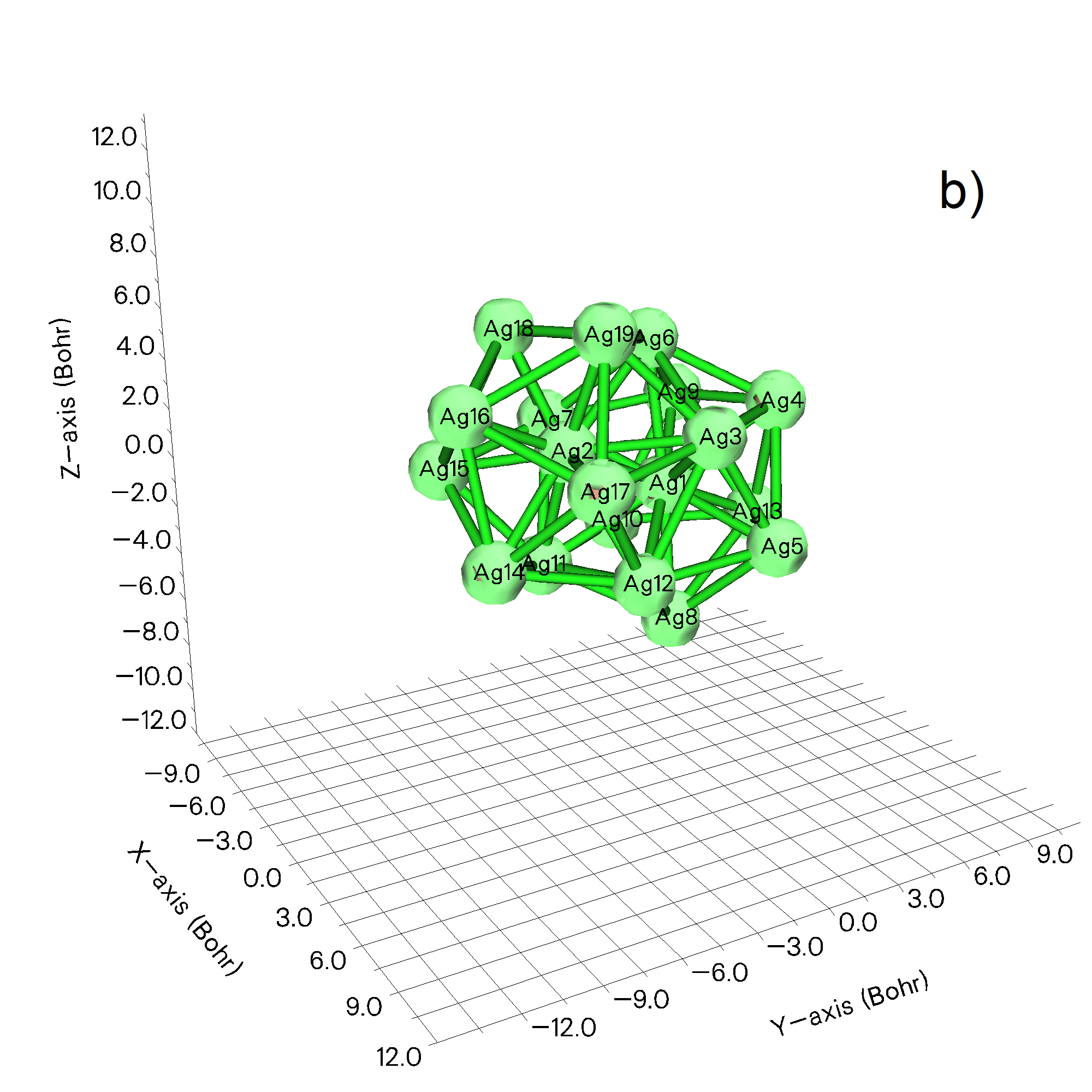}} 
\end{figure}

To investigate the stability and reactivity of double icosahedron Ag$_{17}$M$_2$ (M = Ni, Cu, Zn) clusters, we summarized the calculated parameters in Table~\ref{table1}. The results show that doping the Ag$_{19}$ cluster increases the ionization energy, while the electron affinity does not show a specific trend upon doping. The derived parameters such as the chemical hardness ($\eta$) and electronegativity ($\chi$) show increased values for the doped clusters compared to the bare Ag$_{19}$ cluster, suggesting that the doped clusters are more prone to accepting electrons. 

\begin{table}[ht!]
{
\caption{\label{table_2}{The ionization energy (I), electron affinity (A), chemical hardness ($\eta$), electronegativity ($\chi$) and electrophilicity index ($\omega$) of Ag$_{17}$M$_2$ (M = Ni, Cu, Zn) and Ag$_{19}$ clusters. The results are calculated with the PB86 functional in conjunction with the Def2-ECP basis set. The energy is given in eV.}}
\small
\def\arraystretch{1.1}
\begin{tabular}{p{1.8cm}p{1.2cm}p{1.2cm}p{1.2cm}p{1.2cm}p{1.2cm}p{1.2cm}}
{Cluster}   &  I     & A    & $\eta$ & $\chi$  & $\omega$  \\ \hline
Ag$_{17}$Ni$_{2}$ &  5.67   & 2.40  &  1.63  &  4.04 & 5.00 \\  
Ag$_{17}$Cu$_{2}$ &  5.45   & 2.48  &  1.48  &  3.96 & 5.29  \\
Ag$_{17}$Zn$_{2}$ &  5.76   & 2.55  &  1.60  &  4.15 & 5.39 \\
\\
Ag$_{19}$          &  5.36   & 2.45  &  1.45  &  3.91 & 5.26 \\
\hline
\end{tabular}
	}
\end{table}

To gain insight into the reactivity of the clusters, we evaluated the ELF distribution, which indicates where electrons are localized and helps identify active sites that are either electron-rich (nucleophilic) or electron-deficient (electrophilic). These sites may correspond to bonding interactions between atoms or molecules with the cluster. In a metal nanoparticle catalyst, it is well established that atoms at edges and corners exhibit higher reactivity due to their lower coordination numbers.\cite{Zhang2017,doi:10.1021/nl2017459} This increased reactivity is associated with enhanced electron localization, which facilitates interactions with reactants. We expect high ELF values to indicate regions where electrons are highly localized and likely to interact with reactants. As shown in Figure~\ref{figura4}, the doped clusters, for example, Ag$_{17}$Ni$_2$, exhibit more localized sites on the Ag surfaces but greater depletion at the central sites, which can be contrasted with the bare Ag$_{19}$ cluster. These preliminary results provide a foundation for further exploration of cluster reactivity, with potential applications in catalysis.
\cite{10.1063/1.4935566,D2CP05188E,RODRIGUEZKESSLER201820636,RODRIGUEZCARRERA2024122301,OLALDELOPEZ2024,doi:10.1021/acs.jpcc.8b09811} \\

\section{Conclusions}

In this preprint, the structural and electronic properties of Ag$_{17}$M$_2$ (M = Ni, Cu, Zn) clusters are investigated using density functional theory (DFT) calculations. The DFT analysis of the clusters confirms their preference for endohedral configurations in the doublet state across multiple functionals. Doping enhances the ionization energy and electronegativity of the clusters compared to the bare Ag$_{19}$ counterpart, indicating increased stability. Furthermore, the observed increase in ELF values at Ag sites suggests enhanced electron localization, which may have significant implications for catalysis.


\section{Acknowledgments}
P.L.R.-K. would like to thank the support of CIMAT Supercomputing Laboratories of Guanajuato and Puerto Interior.



\bibliographystyle{unsrt}
\bibliography{mendeley}
\end{document}